\documentclass[twocolumn]{article}
\usepackage{graphicx}
\makeatletter
\def\@normalsize{\@setsize\normalsize{10pt}\xpt\@xpt
\abovedisplayskip 10pt plus2pt minus5pt\belowdisplayskip \abovedisplayskip
\abovedisplayshortskip \z@ plus3pt\belowdisplayshortskip 6pt plus3pt
minus3pt\let\@listi\@listI}
\def\subsize{\@setsize\subsize{12pt}\xipt\@xipt}
\def\section{\@startsection {section}{1}{\z@}{1.0ex plus 1ex minus
 .2ex}{.2ex plus .2ex}{\large\bf}}
\def\subsection{\@startsection {subsection}{2}{\z@}{.2ex plus 1ex}
{.2ex plus .2ex}{\subsize\bf}}
\makeatother
\begin{document}
\date{}
\title{\bf The rules of coherence and other habits}
\author{M. R. C. Solis\\
  Theoretical Physics Group \\
  National Institute of Physics \\
  University of the Philippines, \\
  Diliman,Quezon City\\
  1101 PHILIPPINES\\
  e-mail: msolis@nip.upd.edu.ph}
\maketitle
Teachers and students alike have often wondered why some people are more  
succesful at mathematics and physics than others. It is tempting to believe that some people 
are just smarter than others. It is a flattering explanation, if you happen to be among the 
succesful people, while at the same time offering consolation for other people, e.g. ``I just don't have 
a mathematical mind''. 

I believe though, that there is another explanation. People good at math and physics are 
succesful not because they're smarter than the rest of us (although being a bit smarter than 
the rest of us does help!). Rather, the explanation is found in the habits that they've 
developed.

It is unfortunate that these habits aren't often taught to freshmen because they stand to 
benefit most. It is hoped that some effort be made in physics departments to promote these 
habits. Oxford and Cambridge University provide booklets explaining these points to their 
freshmen mathematics majors. Something similar might be done for freshmen physics majors.

To explain these habits, I shall concentrate on how to apply these habits to the study of 
mathematics. This article contains some observations on 
mathematically successful people as well as some of the tricks that I have learned while 
grappling with the material. This article doesn't contain any new mathematics; instead, it is 
intended to help you learn how to do mathematics.\footnote{The methods suggested here 
are also useful in physics!} 

\section{The Rules of Coherence}

In this section, we will learn how to write down solutions. This is an important skill, 
because continued practice of the rules of coherence forces you to seek the clearest 
explanation. Furthermore, a clear explanation is easier to remember than a complicated one. 

The rules of coherence also help you write papers that you intend to submit for publication.
By following these rules, you may go a long way towards making the reading of your work a 
pleasurable experience.

\begin{itemize}

\item {\bf Math is prose.} This means mathematics (at least, as used by the physicist) is a 
language as well as a means towards getting the answers we want. 
Since it is a language, we must make an effort to integrate it carefully with our ordinary 
English prose, or whatever language we are using. For example,  ``using the angle addition 
formulae for cosines
\begin{eqnarray}
\mbox{cos(x+y)}=\mbox{cos}x\mbox{cos}y-\mbox{sin}x\mbox{sin}y \\
\mbox{cos(x-y)}=\mbox{cos}x\mbox{cos}y+\mbox{sin}x\mbox{sin}y \label{addition}
\end{eqnarray}
we may easily derive the formula...''

This is an important rule that most students seem to forget whenever they have to pass 
solutions to assigned problem sets. What they submit is usually a stream of equations without 
signposts for the poor reader (in this case, the instructor) as to how they got which from 
what. Symbols are left undefined, assumptions and theorems used are not explicitly stated, 
etc. Is it any wonder that the instructor therefore descends like an avenging angel on their 
work, trailing red ink in his wake?

\item {\bf Fisher's rule} This rule is named after one of the teachers of N. David Mermin.
It is simply stated: Number {\em all} equations. We do this to make our solutions clear. 
Instead of referring to ``the first equation in the introduction,'' we instead say something 
like ``from the angle addition formulae for cosines (\ref{addition}).'' 

There is a variant of Fisher's rule, called Occam's rule, which Mermin calls a heresy: Number 
all equations you think you might want to refer to. I do not recommend it. For example, what 
happens if you wish to add another equation, or what if you eventually want to cite an 
equation that you left unnumbered? You would have to go through the painful process of 
renumbering by hand. This is okay if your work is a mere three pages, but it is a nightmare 
if you wish to fix a two hundred-page manuscript.

\item {\bf The Good Samaritan Rule} A good Samaritan is someone who helps people in distress, 
and nothing is more distressing than the appearance of ``from equation (3.3.10)'', an 
equation found in chapter3, when you are working on something in, say, chapter 17. It is 
annoying especially when you're reading something as massive as MTW{\footnote {{\bf 
Gravitation} by Misner, Thorne, and Wheeler (a classic General Relativity text )}}, and you 
have to flip through five hundred pages just to figure out what the equation is all about. 

Thus, Mermin proposed the Good Samaritan rule: Add explanatory text. For example, when citing 
an equation, introduce it with a descriptive phrase as well as an equation number.

You might ask, ``Why bother with the good samaritan rule at all?''. After all, you might 
think, you understand your solution. It may be so today, but what if you had to read your 
solution three years from now? (See ``file your work'') The good samaritan rule exists so that 
you will be able to understand your solution, say, twenty years from now. 

The good samaritan rule is also useful when you're writing computer programs. It isn't hard to 
add explanatory text (comments!) to computer programs. Comments make the job of understanding 
your code easier for other people and yourself.

\item {\bf Cite your sources.} This rule is about honesty. If your pal helped you work out a 
difficult mathematical problem, acknowledge his or her help. If your formula came from Arfken 
or from Whittaker and Watson, say so and include the page (and equation number), edition, 
etc.

Aside from honesty, citing sources is especially useful if you want to recheck your work. 
Sources aren't infallible. Reproducing previously done work adds to the 
length of our soultions. Citing sources is a convenient way of avoiding having to redo what 
might be a long calculation. \footnote{Although it doesn't remove the burden of making sure 
that the source we used is indeed reliable!}
\end{itemize}

The rules of coherence boils down to a simple question: can my solution serve as supplementary 
lecture notes? If yes, then the rules of coherence are satisfied. If no, you need to rework 
your solution until it satisfies the rules of coherence.\footnote{This statement of the rules of coherence is due to
N. David Mermin, unpublished lecture notes.}

\section{Other habits}

Aside from the rules of coherence, there are other useful habits.

\begin{itemize}
\item {\bf File your work.} It is part of the fallen state of man that he forgets. If you 
spent a week thinking about a problem, and you were asked about it three years later, 
wouldn't it be frustrating to find out that you have {\em absolutely} forgotten {\em how} to 
solve the problem? If you had it on file, re-reading your work could save you what might have 
been another week of thinking.

There is no guarantee that the solutions you submit will be returned to you. Therefore, 
always make two copies of your solution-- one for yourself, and another for your instructor.
This isn't difficult today, for ours is the age of the photocopying machine.
 
This is how I file my work. I have folders with labels for the various things that interest 
me. As soon as I have solved something, I produce a neat write-up and file it in the relevant 
folder. I make sure that I write down when I solved it, and if the problem was really tough, 
I also include wrong attacks on the problem, as well as the length of time (days, weeks, 
months, or even years) I spent on it. You might have your own filing system. If so, use the 
filing system that is most convenient for you. So long as you file whatever it is that you 
worked on.

\item {\bf Study ahead.} This is something that I discovered after carefully watching the 
ur-Geek in the theoretical physics group. Before taking a graduate-level quantum mechanics 
course, this guy sat through a year of lectures (as an undergrad!) and worked on a sizable 
number of problems. Is it any wonder he aced the class?

\item {\bf Solve as many problems as you can.} The only way to test your understanding of the 
material in the lectures or in your book is by working on the problems. I am of course, 
assuming that you follow the other rules as well. The aforementioned rules help keep you 
honest. It is almost impossible to fool yourself if you conscientiously follow Fisher, the 
Good Samaritan, etc. If there are gaps in your understanding, you will eventually reach a 
point where you cannot justify the steps you made. When you find yourself unable to explain 
the origins of an equation, it means you have to do some more reading and thinking.

\item {\bf Talk to other people.} One of the best ways of learning is by talking it over with a friend, a colleague, or 
a professor. Try to explain the idea you've learned to other people. It is very easy to fool oneself. Other people 
provide feedback by asking questions and making you clarify things.
\end{itemize}  

Incidentally, the rules of coherence and the other habits might provide an explanation for the 
Feynman effect. As everyone knows, Feynman is legendary for his deep understanding of physics. 
Based on secondhand reports \footnote{James Gleick, {\em{Genius: The Life and Science of Richard Feynman
}}, Vintage Books; Reprint edition (November 1993) 
}, Feynman had an extensive filing syste. Feynman even preserved his Far Rockaway notebooks, 
the ones he used as a high school student to teach himself trigonometry and calculus! By 
filing his work, he had an enormous storehouse of previous work, just waiting for an 
opportunity to be used in other areas of his research.

\section{Final Words}
In this article, we have examined some helpful habits. There are others, of course.
Based on an informal survey  these habits aren't as widespread among undergraduate physicists. 
This is a pity, because we have a lot to gain from these habits. 

Physics and mathematics are difficult enough without the aditional burden of bad habits. 
It is hoped that by explaining these habits to our students, and encouraging the adoption of 
these habits, they will end up with a deeper understanding of mathematics and physics. 

\subsection*{Acknowledgements}
The author wishes to thank the members of the GR study group: A. Alarcon, B. A. Rara, G. M. Sardane, 
M. S. A. Sereno, and R. R. D. J. Sol. He also wishes to thank N. D. Mermin and E. F. Taylor for reading 
the manuscript, and for their comments and suggestions.

\end{document}